\documentclass[english,aps, prl,reprint,twocolumn, twoside,showkeys]{revtex4-1}
\usepackage[T1]{fontenc}                                                                  
\usepackage[latin9]{inputenc}                                                             
\usepackage{graphicx}                                                                     
\usepackage{amsmath}          
\usepackage{amssymb}                                                                     
\usepackage{rotating}
\usepackage[colorinlistoftodos]{todonotes}
\usepackage[centering,hmargin=2.5cm,vmargin=2.5cm]{geometry}

\newcommand{\Ref}[1]{Ref.~\onlinecite{#1}}

                                                                                          
                                                                                          
\usepackage{babel} 
\begin{document}  
                             
\title{Many Molecular Properties from One Kernel in Chemical Space} 
\date{\today}     

\author{Raghunathan Ramakrishnan$^{1}$}
\author{O. Anatole von Lilienfeld$^{1,2}$}
\email{anatole.vonlilienfeld@unibas.ch}
\affiliation{$^1$Institute of Physical Chemistry and National Center for Computational Design and Discovery of Novel Materials (MARVEL), Department of Chemistry, University of Basel, Klingelbergstrasse 80, CH-4056 Basel, Switzerland}
\affiliation{$^2$Argonne Leadership Computing Facility, Argonne National Laboratory, 9700 S. Cass Avenue, Lemont, IL 60439, USA}

\begin{abstract}
We introduce property-independent kernels for 
machine learning modeling of arbitrarily many molecular properties.
The kernels encode molecular structures for training sets of varying size,
as well as similarity measures sufficiently diffuse in
chemical space to sample over all training molecules.
Corresponding molecular reference properties provided, they 
enable the instantaneous generation of ML models which can systematically be
improved through the addition of more data.
This idea is exemplified for single kernel based modeling of
internal energy, enthalpy, free energy, heat capacity,
polarizability, electronic spread,
zero-point vibrational energy, 
energies of frontier orbitals, 
HOMO-LUMO gap,
and the highest fundamental vibrational wavenumber.
Models of these properties are trained and tested using 112 kilo organic molecules of similar size. 
Resulting models are discussed as well as the kernels' use 
for generating and using other property models.

\end{abstract}

\maketitle

\section{Introduction}
Strategies for solving computational chemistry problems have evolved in parallel with the capacity and abundance of  computer hardware \cite{wilson2000chemists}. 
Access to ever-increasing compute power available at centralized computer facilities, such as the sciCORE at the University of Basel, the Swiss National Supercomputing Centre or the Argonne Leadership Computing Facility, enable ``Big Data'' driven computational chemistry which no longer relies on experimental data for training and validation, but rather
on virtual data, obtained through predictive modeling and massive simulation efforts.  
Statistical inference from Big Data holds great promise in many scientific domains,
including biology \cite{marx2013biology}, 
climate research \cite{mattmann2013computing},
high-energy physics \cite{doctorow2008big}, 
or photonics \cite{wright2014big}. 
Due to their unrivaled computational efficiency 
(typical execution time is milliseconds),
data driven models of molecular property predictions become relevant as soon as they
reach the accuracy of well established deductive quantum chemistry methods for
solving approximations to the electronic Schr\"odinger equation, such as
Hartree-Fock, Density Functional Theory (DFT), or Coupled-Cluster methods.

For the ``supervised learning'' task \cite{mohri2012foundations} of inferring 
a molecular property from structure-property dyads provided {\em a priori}, 
machine learning (ML) algorithms have very recently been shown to
reach desirable quantum chemical accuracy, even when predicting 
properties for new (out-of-sample) molecules which had no part in training~\cite{rupp2012fast,montavon2012learning,von2013first,montavon2013machine,hansen2013assessment}.
These developments have also triggered studies on
transition state dividing surfaces~\cite{ML4Graeme2012},
orbital-free kinetic density functionals~\cite{ML4Kieron2012},
electronic properties of crystals~\cite{GrossMLCrystals2014},
transmission coefficients in nano-ribbon models~\cite{lopez2014modeling},
or densities of states in Anderson impurity models~\cite{arsenault2014machine}.
In order to establish a consistent dataset for which ML models
can be improved systematically through the addition of more data, 
we recently published computed DFT structures and multiple properties of 134 thousand (134\,k) organic molecules~\cite{ramakrishnan2014quantum}. 
Within a previous study \cite{ramakrishnan201bigdata},
discussed at the Swiss Chemical Society Fall Meeting 2014,
we used some properties of this dataset to investigate the
$\Delta$-ML approach that augments less expensive deductive baseline theories 
by inductive ML models, trained on the baseline's deficiencies.
We demonstrated that chemical accuracy can be reached within this $\Delta$-ML approach, 
as well as strong transferability when applied to all the 134\,k molecules. 
In this study, we investigate the use of property-independent kernels
for the simultaneous modeling of multiple properties taken from the
same data base~\cite{ramakrishnan2014quantum}. 
To this end, we rely on kernel functions sufficiently diffuse
to account for significant similarity measures among all training molecules. 
This enables us to reapply inverted kernel matrices to any arbitrary
set of molecular properties and to generate the corresponding ML models
on the same footing.
We have validated our approach by simultaneous training and 
prediction of 13 energetic and electronic molecular properties.

This article is organized as follows:
Section 2 Methods briefly summarizes the ML notations and definitions, along with 
a discussion of kernel  function shape and spread. 
In Section 3 Computational Details we describe our molecular data selection strategy, 
and discuss the selection of properties in the dataset. 
We present and analyze our results for the
performance of property-independent kernels in Section 4.
In Section 5 we draw our conclusions. 
In the appendix we explain how to access and reuse the kernel data.
  

\section{Methods}
Inarguably, one of the more appealing ML algorithms is kernel-ridge-regression (KRR) \cite{kung2014kernel} because
of its numerical robustness and conceptual simplicity.
Within KRR, the ML {\it Ansatz} for a given property $p$ of any 
query molecule, $q$, is merely a linear combination of similarity measures 
between $q$ and a finite set of $N$ training molecules $t$,
\begin{eqnarray}
p_q = \sum_{t=1}^{N} c_t^p K_{qt}.
\label{eq:ansatz}
\end{eqnarray}
In Eq.~(\ref{eq:ansatz}) $K_{qt}$ is a kernel matrix element corresponding to molecules $q$ and $t$. 
Here, we wish to investigate if $K_{qt}$ can be made independent of $p$.
We have chosen to use the exponentially decaying (a.k.a. Laplacian) 
kernel function of similarity measure between $t$ and $q$,
$K_{qt}=\exp\left(-D_{tq}/\sigma\right)$, where $D_{tq}=|{\bf d}_t - {\bf d}_q|$ is the Manhattan (a.k.a. $L_1$)
norm of difference between two, typically non-scalar, 
descriptors of molecules $t$ and $q$, respectively. The global 
hyperparameter $\sigma$ quantifies the kernel width \cite{kung2014kernel}.
For the Coulomb matrix (CM) descriptor, the combination of Laplacian kernel with $L_1$ norm
has been shown to yield good ML models of atomization energies
\cite{hansen2013assessment}.

Prior to predicting molecule $q$'s property $p_q$ according to Eq.~(\ref{eq:ansatz}), 
the vector ${\bf c}^p$, with optimal regression coefficient $c_t^p$ as element for every training molecule $t$,
must be obtained through minimization of the penalized Lagrangian function,
\begin{eqnarray}
\mathcal{L}=\left( {\bf p}^{r} - {\bf K}{{\bf c}^p} \right)^{\rm T} \left( {\bf p}^{r} - {\bf K}{{\bf c}^p} \right) +\lambda {{\bf c}^p}^{\rm T}{\bf K}{{\bf c}^p},
\label{eq:lag}
\end{eqnarray}
where matrices are in upper cases, vectors in lower case, and $()^{\rm T}$ denotes a transpose.
${\bf p}^{r}$ represents the vector with reference property values for all
$N$ training molecules, and ${\bf K}$ is the kernel matrix with above mentioned
elements.
The $\lambda$-term imposes regularization while the first term 
corresponds to the conventional least square regression.
Setting the derivative of $\mathcal{L}$, with respect to ${{\bf c}^p}$, to zero, 
the coefficients which minimize Eq.~(\ref{eq:lag}) can be shown to amount to
\begin{eqnarray}
{{\bf c}^p} = \left( {\bf K} + \lambda {\bf I}\right)^{-1} {\bf p}^{r}.
\label{eq:inv}
\end{eqnarray}

The dependence of a model's performance on hyperparameters, $\sigma$, and $\lambda$, can be understood as follows. In the presence of training molecules with extremely outlying properties an optimal value of $\lambda$ becomes non-zero in order to quench excessively large elements of ${{\bf c}^p}$. 
In other words, the modeling function becomes more rigid and lessens the danger of overfitting.
The meaning of the kernel width, $\sigma$, is to control the ${c^p_t}$ contribution from training molecule $t$ when making a new prediction, see Eq.~(\ref{eq:ansatz}). 
Typically, optimal choices of $\sigma$, and $\lambda$ are obtained for every molecular property and trainingset size
through extensive use of cross-validation (CV) within training molecules.
CV can become a computational bottleneck for larger training sets. 
For example, for training sets of $N$ = 10\,k, a 5-fold CV implies to repeatedly invert 8\,k$\times$8\,k
matrices, each requiring $\sim$0.5 CPU hours on modern compute hardware. 

In this study, we have explored the possibility to always keep all training molecules fixed,
to estimate the hyperparameter $\sigma$ beforehand, and to set $\lambda$ to zero.  
This allows us to obviate all CVs, and to use a single identical kernel matrix ${\bf K}$ for any properties. 
More specifically, once ${\bf K}^{-1}$ has been computed and stored,
regression coefficients for any number of properties, $p_1, p_2, ..., p_n$, 
can be computed simultaneously, 
\begin{eqnarray}
\left[ {\bf c}^{p_1} {\bf c}^{p_2} \ldots {\bf c}^{p_n} \right] & =& 
{\bf K}^{-1} \left[ {\bf p}^{r}_1 {\bf p}^{r}_2 \ldots {\bf p}^{r}_n \right] \quad
\Rightarrow \quad {\bf C}={\bf K}^{-1} {\bf P}^{r},\nonumber \\
\label{eq:inv2}
\end{eqnarray}
where $\bf{C}$ and ${\bf P}^{r}$ are the $N\times n$ regression coefficient and property matrices, respectively. 
Consequently, instead of  $n$ CVs 
with computationally demanding multiple kernel matrix inversions, 
each scaling as ${\mathcal O}(N^3)$, 
the computational cost is now being dominated by one kernel inversion plus $n$ matrix-vector multiplications, each scaling as ${\mathcal O}(N^2)$.

\begin{figure*}[hpt]
\centering      
\includegraphics[width=12.3cm, angle=0.0, scale=1]{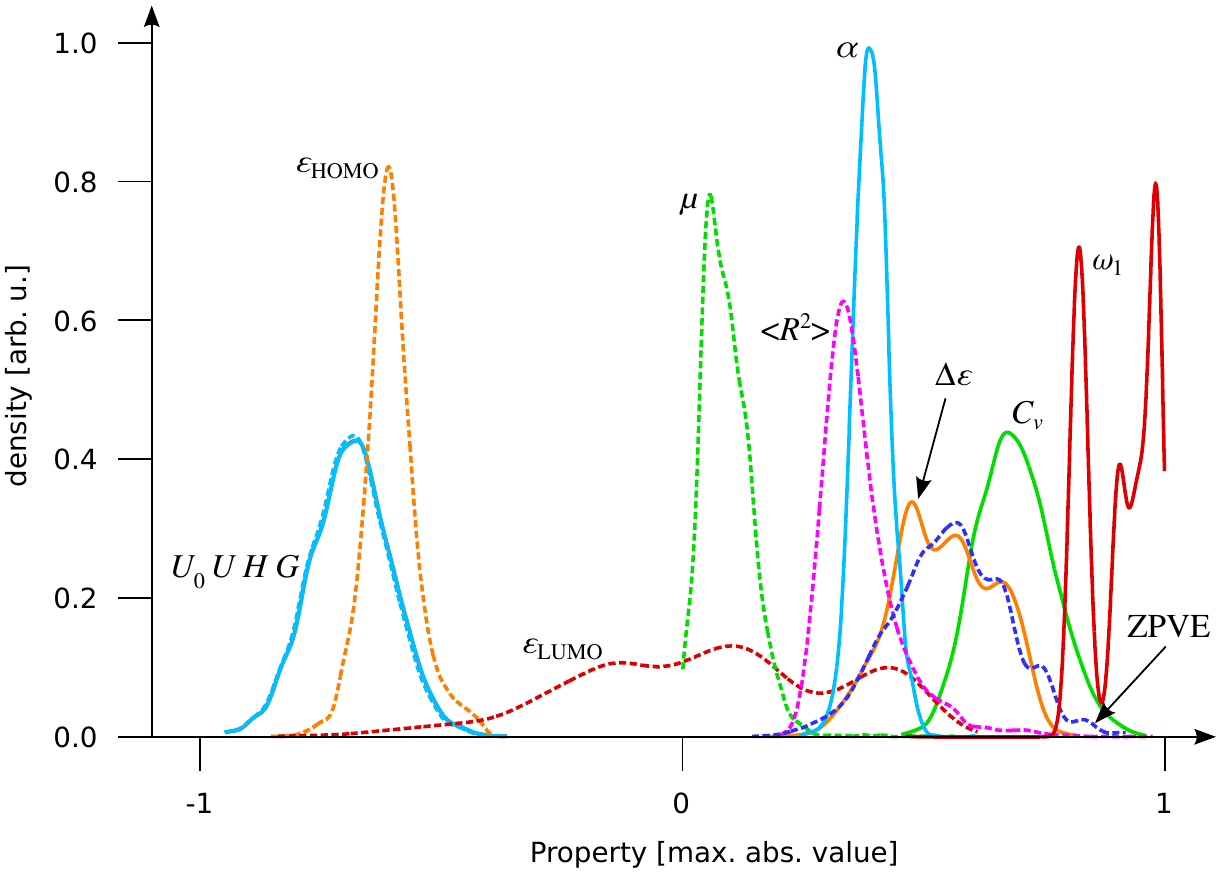}                              
\caption{Density distributions of thirteen properties for the set of 112\,k  organic molecules made up of CHONF. 
The abscissa corresponds to values of the properties
relative to corresponding maximal absolute values in the dataset: 
$|U_{0}|_{\rm max}=2608.5$ kcal/mol,
$|U|_{\rm max}=2626.4$ kcal/mol,
$|H|_{\rm max}=2643.0$ kcal/mol,
$|G|_{\rm max}=2417.1$ kcal/mol,
$|\alpha|_{\rm max}=196.6~a_0^3$,
$|C_{v}|_{\rm max}=$47.0 cal/mol/K,
$|\mu|_{\rm max}=30.0$ D,
$|\langle R^2 \rangle|_{\rm max}=3375~a_0^2$,
|ZPVE|$_{\rm max}$=171.9 kcal/mol,
$|\varepsilon_{\rm HOMO}|_{\rm max}$=10.78 eV,
$|\varepsilon_{\rm LUMO}|_{\rm max}$=4.68 eV,
$|\Delta \varepsilon|_{\rm max}$=12.57 eV, and
$|\omega_1|_{\rm max}$=3876.7 cm$^{-1}$.
}
\label{fig:distr}
\end{figure*}

\section{Computational Details}

For typical molecular datasets, with $N>5$\,k, we find that optimal $\sigma$ converge towards a large value which ensures non-vanishing contributions from all training compounds, while $\lambda$ converges towards zero. 
The choice of setting $\lambda$ to zero is easily justified: 
As also seen below, property values in the molecular dataset show distributions
centered at an average value with relatively few outliers, if any.
Over-fitting, leading to large coefficients for these outliers, is therefore unlikely to influence
the performance of ML models based on thousands of well-behaved training molecules. 
For the Laplacian kernel used here, extreme (too large/small)
values of  $\sigma$ lead to loss of information regarding  
training descriptor distances, $D_{ij}$. 
For  $\sigma \approx 0$, off-diagonal elements of the kernel matrix, $K_{ij}=\exp\left(-D_{ij}/\sigma\right)$
vanish, resulting in a unit-kernel-matrix, ${\bf K}={\bf I}$. 
On the other hand, for $\sigma >> 1$, a kernel-matrix of ones is obtained which would be
singular (and hence non-invertible) for $N>1$, and anyways does not resolve $D_{ij}$. 
Here, we define the optimal $\sigma$ value to be defined such that all kernel elements
are in between 0.5 and 1. This can be accomplished
through the constraint for the smallest kernel matrix element, 
i.e. the kernel element which corresponds to the two most distant training molecules, 
\begin{eqnarray}
\exp\left(-D_{ij}^{\rm max}/\sigma_{\rm opt}\right) = 1/2,
\label{eq:optsigma}
\end{eqnarray}
resulting in 
$\sigma_{\rm opt} = D_{ij}^{\rm max} / \log(2)$.
As such, through use of $\sigma_{\rm opt}$ and $\lambda =$ 0, a property-independent global 
kernel matrix is obtained which only needs to be inverted once before it is used
to generate regression coefficients for all molecular properties of interest. 
For randomly sampled 1\,k molecules, 
from the dataset considered in this study ({\it vide infra}), $D_{ij}^{\rm max}=677$ a.u., suggesting
$\sigma_{\rm opt} \approx 977$ a.u. 
This number is consistent with a numerical grid search for $\sigma_{\rm opt}$ which
identifies $\sigma =$ 1000 a.u. to be optimal for a 
1\,k ML model of atomization enthalpies. 
In the remainder of this study we discuss the performance of ML models based on inverted
global kernels for up to 13 molecular properties, always with $\sigma_{\rm opt}=1000$ a.u.
and $\lambda =$ 0, irrespective of the training set size $N$.

We have also investigated if our observations depend on the choice
of molecular descriptor, ${\bf d}$. To this end, we have considered 
two different descriptors, namely the Coulomb-matrix (CM) with 
rows and columns uniquely permuted \cite{von2013first}, 
as well as the bag-of-bonds (BOB) \cite{BobPaper} descriptor,
amounting to an ordered set of weighted interatomic distances.

We have used the published \cite{ramakrishnan2014quantum} quantum chemistry results
for the smallest 133885 (134\,k) organic molecules subset of the GDB-17 published by Reymond {\it et al.} \cite{ruddigkeit2012enumeration}, which contains over 166\,giga molecules.
This 134\,k dataset contains relaxed geometries and chemical properties computed using the
DFT (B3LYP with basis set 6-31G(2df,p)). Here, we have pruned this dataset by
eliminating all molecules with up to 8 ``heavy'' atoms (not counting hydrogens) which, trivially, would be outliers since the dataset is dominated by molecules with 9 heavy atoms. 
For the resulting 111594 (112\,k) molecules, 
we have considered the following 13 computed properties: Zero-Kelvin internal energy $U_0$,
thermochemistry energetics at 298.15\,K (internal energy $U$, enthalpy $H$, free energy $G$,
all three properties for the process of atomization, and
heat capacity $C_v$); 
isotropic molecular polarizability, $\alpha$; 
electronic radial expectation value, $\langle R^2 \rangle$; 
harmonic zero-point vibrational energy, ZPVE; 
energy of highest  occupied molecular orbital (HOMO), $\varepsilon_{\rm HOMO}$; 
energy of lowest unoccupied molecular orbital (LUMO), $\varepsilon_{\rm LUMO}$; 
HOMO-LUMO gap, $\Delta \varepsilon$; 
and the highest fundamental vibrational wavenumber, $\omega_1$, 
in the 3000--3900 cm$^{-1}$ range.

Fig.~\ref{fig:distr} features the 
density distributions of the relative values of these properties, scaled 
with respect to their maximal values in the dataset. 
We note the properties with large property density, 
polarizability, dipole moment, energy of HOMO,
and radial expectation value. Compared to these properties, 
all thermochemical properties 
are less densely distributed. 
Properties even more sparsely distributed include
LUMO energy, gap and ZVE; interestingly their densities also
exhibit multimodal distributions (see Fig.~\ref{fig:distr}),
possibly arising from characteristic functional group moieties present in the dataset.
The distribution of $\omega_1$ shows three narrow peaks, which can readily be interpreted  
as arising from C-H, N-H, and O-H (symmetric and asymmetric) stretching modes.
Corresponding values from literature 
\cite{johnson2013nist} for similar wavenumbers at the same level of theory read
for CH$_4$ (${\rm A_1}$, 3038, and ${\rm T_2}$, 3152 cm$^{-1}$),
NH$_3$ (${\rm A_1}$, 3459, and ${\rm E}$, 3576 cm$^{-1}$),
and H$_2$O (${\rm A_1}$, 3802, and ${\rm B_2}$, 3906 cm$^{-1}$).
Further details regarding the genesis of the dataset can be found in
\Ref{ramakrishnan2014quantum}.

\section{Results and discussion}

\begin{figure}[hpt]
\centering      
\includegraphics[width=5.8cm, angle=0.0, scale=1]{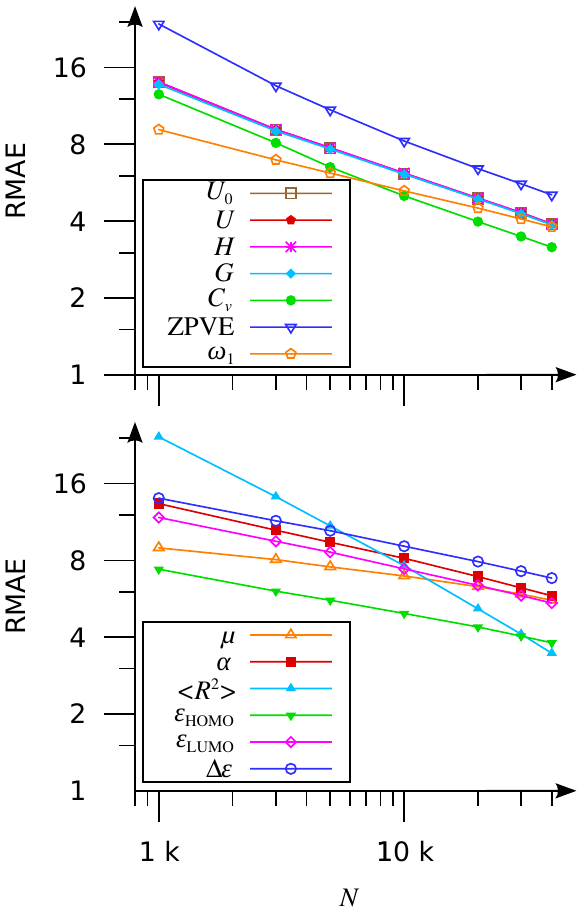}   
\caption{
Relative mean absolute errors (RMAE) in all ML predicted properties of out-of-sample
molecules, in the 112\,k set using the same kernel function of size up to $N=40$\,k.
See text for the definition of RMAE for respective properties. 
TOP: Thermochemistry and vibrational properties. BOTTOM: Electronic properties.
}
\label{fig:LC}
\end{figure}

Using single kernels of varying size, the systematic decay of ML prediction errors 
is summarized in Fig.~\ref{fig:LC} for all the aforementioned properties.
Note that all reported error measures refer to out-of-sample predictions, 
i.e. for a given training set of size $N$, errors are presented as measured
on the remaining 112\,k - $N$ molecules.
To compare the error across different properties, irrespective of units and scale,
we have used the mean absolute error (MAE)
relative (i.e., RMAE) to desired quantum chemistry accuracy norms 
as a suitable error measure. The target accuracy for the thermochemical quantities,
and  orbital energies is 
the highly coveted ``chemical accuracy'' for energetics, i.e., 1 kcal/mol. For
$\omega_1$, and ZPVE, both 
within the harmonic approximation, we have selected a target accuracy of
10~cm$^{-1}$. 
This value is slightly larger than the average accuracy of  coupled cluster method, CCSD(T) + with converged basis sets \cite{tew2007basis}, for predicting harmonic wavenumbers of small molecules, 
as measured by comparison to their experimentally determined counterparts. 
For dipole moment and isotropic polarizability, the target accuracies employed are 0.1 D,
and 0.1 $a_0^3$, respectively. These thresholds are within the uncertainty of
predicted values of these properties at the CCSD level of theory \cite{hickey2014benchmarking}.

The most compelling feature in Fig.~\ref{fig:LC} is the 
systematic decay in RMAEs for {\it all} molecular properties. 
This amounts to numerical evidence that predictive ML-models for multiple properties can be built
using a single kernel matrix with no property-specific parametrization.  
Furthermore, the accuracy of these single kernel models can systematically
be improved through the addition of more training data.
Among all properties, we note maximal learning rates for
$\langle R^2 \rangle$---a measure of
diffuseness of the electron density, and quite possibly
more directly linked to molecular geometry and
composition than the other observables.
ZPVE exhibits the second best learning rate,
implying a potential scope to invest on building larger ML models for the 
data-driven, rapid and accurate estimations of 
ZPVE corrections for a multitude of energetics such as 
barrier heights, reaction/dissociation energies, and thermochemistry. 
 At the limit of 40\,k training molecules, none of the 13 properties
 were predicted  within an RMAE of 1, indicating the
 need to employ larger training set sizes.
$\mu$ and $\omega_1$ are the more difficult properties to learn as evinced in Fig.~\ref{fig:LC}.
We note that despite their strongly differing distribution
(shown in Fig.~\ref{fig:distr}) HOMO and LUMO values have the same learning rate and
only differ in their off-set. 
For all properties, the near-linear learning curves suggest that
target values of RMAE=1 could be reached if only sufficiently large kernels were constructed.
Due to the logarithmic scaling, however, the necessary computational investment
for training set generation will grow increasingly prohibitive.
We have also found such behavior to be independent of descriptor choice:
When repeating the ML calculations using the new descriptor BOB~\cite{BobPaper},
instead of the Coulomb matrix, we noted the same trends in error decays with
slightly better overall performance.


 \begin{figure}[hpt]
\centering      
\includegraphics[width=5.8cm, angle=0.0, scale=1]{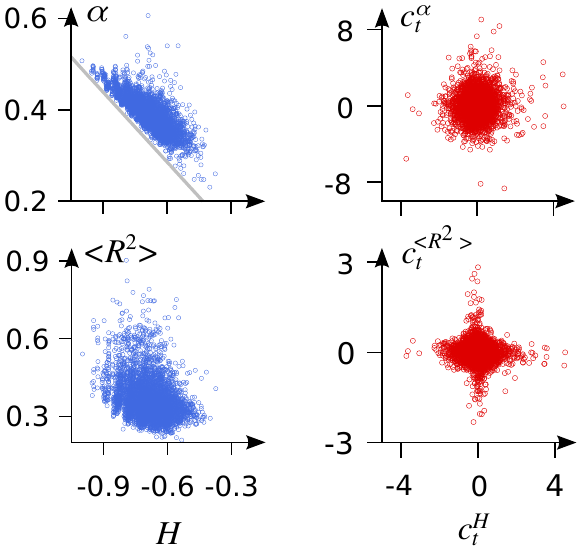}  
\caption{
Scatter plots for properties and corresponding coefficients (using the 40\,k kernel). 
LEFT: Atomization enthalpy, $H$, versus isotropic polarizability, $\alpha$ (TOP), 
and versus radial expectation value, $\langle R^2 \rangle$ (BOTTOM), respectively. 
The gray line ($y=-0.51~x-0.02$) indicates the linear fit to the lower bound $\alpha$ for any given $H$. 
RIGHT: Scatterplots of corresponding regression coefficients, $c^H_t$ vs. $c^\alpha_t$ (TOP),
and $c^H_t$ vs. $c^{\langle R^2 \rangle}_t$ (BOTTOM). 
All values are in units of maximal absolute property as defined in
Fig.~\ref{fig:distr}.
}
\label{fig:sca}
\end{figure}
 
In previous studies we have already analyzed the effect on regression coefficients
of a specific property due to tuning an external parameter~\cite{arsenault2014machine},
here we can exploit the single kernel {\it Ansatz} to directly compare
regression coefficients of different properties for the same training molecules. 
To demonstrate this point we have considered the 40\,k coefficients
used to predict $H$, $\alpha$, and $\langle R^2 \rangle$. 
Fig.~\ref{fig:sca} illustrates the pairwise relationships between properties
$H$ vs. $\alpha$, and $H$ vs. $\langle R^2 \rangle$, as well as
between the corresponding regression coefficients, $c_t$. [Eq.~(\ref{eq:ansatz})]. 
On the one hand, $\alpha$ exhibits the familiar lower linear bound in $H$, 
as also mentioned in \Ref{montavon2012learning},
and reminiscent of the minimal polarizability principle \cite{hohm2000there}, or the related
maximum hardness principle \cite{pearson1993principle}. 
By contrast, the $c^H_t$ vs. $c^\alpha_t$ scatterplot exhibits complete absence
of correlation merely implying a disk shaped bivariate normal distribution 
which sets an upper bound of $(c^H_t)^2$ + $(c^\alpha_t)^2$, for any training molecule, $t$.
On the other hand, and in contrast to  $\alpha$, 
the electron spread, $\langle R^2 \rangle$, correlates rather poorly with $H$ (see bottom panel in Fig.~\ref{fig:sca}). 
The scatterplot for the corresponding $c_t$, however, now shows a characteristic
cross shape, suggesting a mutually complementary mode of action among training molecules. 
More specifically, training molecules very relevant for modeling one property (i.e., large $c_t$)
are insignificant for modeling the other property, and vice versa.
Such a pattern could possibly arise from a bivariate normal distribution bound of
$c^H_t$ and $c^{\langle R^2 \rangle}_t$ with $L_P$-norm for 0 $< P <1$. 

Overall, however, we note that the  distribution of $c_t^p$ 
is governed by the limits imposed through $\sigma_{\rm opt}$. 
For the above mentioned trivial case of ultra-tight kernels, i.e.~in
the limit that $\sigma \rightarrow 0$, 
we have ${\bf K}={\bf K}^{-1}={\bf I}$, and hence ${\bf c}^p={\bf p}^r$. 
This implies that for the diffuse kernel functions
used in this study through choice of 
$\sigma_{\rm opt} = D_{ij}^{\rm max} / \log(2)$, 
one should not expect ${\bf c}$ to reflect the same trends as properties.
Finally we note the coefficients of all properties to show regularized 
distributions that are peaked at zero (i.e. no over-fitting of outliers), 
giving further justification to our choice of $\lambda=0$.

\section{Conclusion}
We have validated and exploited the fact that ML models
based on property-invariant kernels in chemical space 
provide a consistent framework to learn any arbitrary set of global molecular properties 
on the exact same footing. 
Using quantum chemistry data for over 100\,k organic molecules, 
we have presented evidence how this facilitates the simultaneous modeling of several properties.
Numerical results have been discussed for the single kernel based ML model of a wide
variety of electronic and energetic molecular properties, including vibrational wavenumbers. 
Overall, we have generated 182 kernels of varying sizes for this study. 
To enable the reuse of the kernels for new properties by the community, 
we have made them publicly accessible (see Appendix). 
Due to the fact that the computationally 
most demanding step in the development of ML models is matrix inversion, requiring up to 
2 CPU days when using 40\,k training molecules, we include the inverse of the 
kernel matrices along with the data. 
This should enable the accelerated development of models for new properties
that can be applied to any molecules which fall within the $D_{ij}^{\rm max}$ 
of the employed kernel. 
\begin{figure}[hpt]
\centering      
\includegraphics[width=5.8cm, angle=0.0, scale=1]{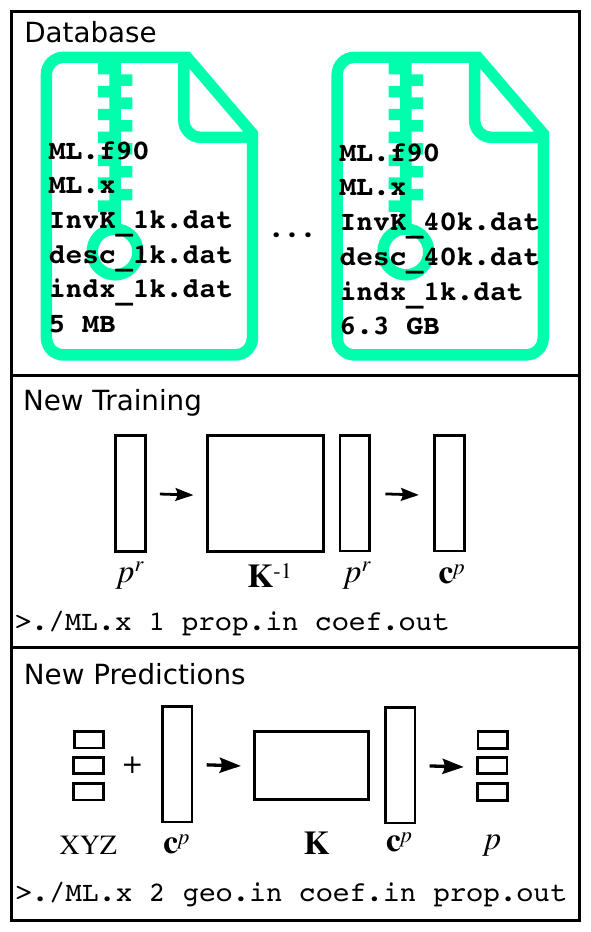}  
\caption{
Schematic description of the database access and usage. 
TOP: For various training set sizes between 1\,k, and 40\,k, 
inverse of kernel matrices, descriptors, an example program, and 
corresponding indices of the training molecules in the 134\,k
dataset \cite{ramakrishnan2014quantum} are archived (see Appendix for details). 
MIDDLE: With the input argument 1, and input property values, the program
calculates the corresponding ${\bf c}$ vector.
BOTTOM: With the input argument 2, ${\bf c}$ vector computed {\it a priori}, and geometries of query molecules, the program estimates the corresponding properties.
}
\label{fig:data}
\end{figure}

\section{Acknowledgements}
OAvL acknowledges funding from the Swiss National Science foundation (No. PP00P2\_138932). 
Some calculations were performed at sciCORE (http://scicore.unibas.ch/) scientific computing
core facility at University of Basel.
This research used resources of the Argonne Leadership Computing Facility at Argonne National Laboratory, which is supported by the Office of Science of the U.S. DOE under contract DE-AC02-06CH11357. 

\section{Appendix: Usage of kernels}

We provide full access to all data generated in this study \cite{datainvk}.
Fig.~\ref{fig:data} illustrates the organization and 
expected usage of the data. 
Using inverse kernel matrices provided in the dataset,
${\bf c}$ vectors (see Eq.~(\ref{eq:ansatz}))
of new properties can be computed through a matrix-vector operation (see Eq.~(\ref{eq:inv})). 
For this purpose, one can use properties reported in \Ref{ramakrishnan2014quantum}, or compute them fresh. It is possible to 
train a model for a property computed at geometries from a theory slightly different than the one employed here. In such cases, 
${\bf c}$ will account for both changes in theory, and in geometries
(see discussions in \Ref{ramakrishnan201bigdata}). This ${\bf c}$ 
vector can then be used for the estimation of 
properties of query molecules with nine of the C, N, O, and F atoms.  
We caution the user that one should not expect predictive power 
for molecules that differ substantially from trainingset molecules in
composition or geometry.

                                                                                          
\bibliography{lit}                                                                 

\begin{thebibliography}{27}%
\makeatletter
\providecommand \@ifxundefined [1]{%
 \@ifx{#1\undefined}
}%
\providecommand \@ifnum [1]{%
 \ifnum #1\expandafter \@firstoftwo
 \else \expandafter \@secondoftwo
 \fi
}%
\providecommand \@ifx [1]{%
 \ifx #1\expandafter \@firstoftwo
 \else \expandafter \@secondoftwo
 \fi
}%
\providecommand \natexlab [1]{#1}%
\providecommand \enquote  [1]{``#1''}%
\providecommand \bibnamefont  [1]{#1}%
\providecommand \bibfnamefont [1]{#1}%
\providecommand \citenamefont [1]{#1}%
\providecommand \href@noop [0]{\@secondoftwo}%
\providecommand \href [0]{\begingroup \@sanitize@url \@href}%
\providecommand \@href[1]{\@@startlink{#1}\@@href}%
\providecommand \@@href[1]{\endgroup#1\@@endlink}%
\providecommand \@sanitize@url [0]{\catcode `\\12\catcode `\$12\catcode
  `\&12\catcode `\#12\catcode `\^12\catcode `\_12\catcode `\%12\relax}%
\providecommand \@@startlink[1]{}%
\providecommand \@@endlink[0]{}%
\providecommand \url  [0]{\begingroup\@sanitize@url \@url }%
\providecommand \@url [1]{\endgroup\@href {#1}{\urlprefix }}%
\providecommand \urlprefix  [0]{URL }%
\providecommand \Eprint [0]{\href }%
\providecommand \doibase [0]{http://dx.doi.org/}%
\providecommand \selectlanguage [0]{\@gobble}%
\providecommand \bibinfo  [0]{\@secondoftwo}%
\providecommand \bibfield  [0]{\@secondoftwo}%
\providecommand \translation [1]{[#1]}%
\providecommand \BibitemOpen [0]{}%
\providecommand \bibitemStop [0]{}%
\providecommand \bibitemNoStop [0]{.\EOS\space}%
\providecommand \EOS [0]{\spacefactor3000\relax}%
\providecommand \BibitemShut  [1]{\csname bibitem#1\endcsname}%
\let\auto@bib@innerbib\@empty
\bibitem [{\citenamefont {Wilson}(2000)}]{wilson2000chemists}%
  \BibitemOpen
  \bibfield  {author} {\bibinfo {author} {\bibfnamefont {E.~K.}\ \bibnamefont
  {Wilson}},\ }\href@noop {} {\bibfield  {journal} {\bibinfo  {journal} {Chem.
  Eng. News}\ }\textbf {\bibinfo {volume} {78}},\ \bibinfo {pages} {39}
  (\bibinfo {year} {2000})}\BibitemShut {NoStop}%
\bibitem [{\citenamefont {Marx}(2013)}]{marx2013biology}%
  \BibitemOpen
  \bibfield  {author} {\bibinfo {author} {\bibfnamefont {V.}~\bibnamefont
  {Marx}},\ }\href@noop {} {\bibfield  {journal} {\bibinfo  {journal} {Nature}\
  }\textbf {\bibinfo {volume} {498}},\ \bibinfo {pages} {255} (\bibinfo {year}
  {2013})}\BibitemShut {NoStop}%
\bibitem [{\citenamefont {Mattmann}(2013)}]{mattmann2013computing}%
  \BibitemOpen
  \bibfield  {author} {\bibinfo {author} {\bibfnamefont {C.~A.}\ \bibnamefont
  {Mattmann}},\ }\href@noop {} {\bibfield  {journal} {\bibinfo  {journal}
  {Nature}\ }\textbf {\bibinfo {volume} {493}},\ \bibinfo {pages} {473}
  (\bibinfo {year} {2013})}\BibitemShut {NoStop}%
\bibitem [{\citenamefont {Doctorow}(2008)}]{doctorow2008big}%
  \BibitemOpen
  \bibfield  {author} {\bibinfo {author} {\bibfnamefont {C.}~\bibnamefont
  {Doctorow}},\ }\href@noop {} {\bibfield  {journal} {\bibinfo  {journal}
  {Nature News}\ }\textbf {\bibinfo {volume} {455}},\ \bibinfo {pages} {16}
  (\bibinfo {year} {2008})}\BibitemShut {NoStop}%
\bibitem [{\citenamefont {Wright}(2014)}]{wright2014big}%
  \BibitemOpen
  \bibfield  {author} {\bibinfo {author} {\bibfnamefont {A.}~\bibnamefont
  {Wright}},\ }\href@noop {} {\bibfield  {journal} {\bibinfo  {journal} {Comm.
  ACM}\ }\textbf {\bibinfo {volume} {57}},\ \bibinfo {pages} {13} (\bibinfo
  {year} {2014})}\BibitemShut {NoStop}%
\bibitem [{\citenamefont {Mohri}\ \emph {et~al.}(2012)\citenamefont {Mohri},
  \citenamefont {Rostamizadeh},\ and\ \citenamefont
  {Talwalkar}}]{mohri2012foundations}%
  \BibitemOpen
  \bibfield  {author} {\bibinfo {author} {\bibfnamefont {M.}~\bibnamefont
  {Mohri}}, \bibinfo {author} {\bibfnamefont {A.}~\bibnamefont {Rostamizadeh}},
  \ and\ \bibinfo {author} {\bibfnamefont {A.}~\bibnamefont {Talwalkar}},\
  }\href@noop {} {\emph {\bibinfo {title} {Foundations of machine learning}}}\
  (\bibinfo  {publisher} {MIT press},\ \bibinfo {year} {2012})\BibitemShut
  {NoStop}%
\bibitem [{\citenamefont {Rupp}\ \emph {et~al.}(2012)\citenamefont {Rupp},
  \citenamefont {Tkatchenko}, \citenamefont {M{\"u}ller},\ and\ \citenamefont
  {von Lilienfeld}}]{rupp2012fast}%
  \BibitemOpen
  \bibfield  {author} {\bibinfo {author} {\bibfnamefont {M.}~\bibnamefont
  {Rupp}}, \bibinfo {author} {\bibfnamefont {A.}~\bibnamefont {Tkatchenko}},
  \bibinfo {author} {\bibfnamefont {K.-R.}\ \bibnamefont {M{\"u}ller}}, \ and\
  \bibinfo {author} {\bibfnamefont {O.~A.}\ \bibnamefont {von Lilienfeld}},\
  }\href@noop {} {\bibfield  {journal} {\bibinfo  {journal} {Phys. Rev. Lett.}\
  }\textbf {\bibinfo {volume} {108}},\ \bibinfo {pages} {058301} (\bibinfo
  {year} {2012})}\BibitemShut {NoStop}%
\bibitem [{\citenamefont {Montavon}\ \emph {et~al.}(2012)\citenamefont
  {Montavon}, \citenamefont {Hansen}, \citenamefont {Fazli}, \citenamefont
  {Rupp}, \citenamefont {Biegler}, \citenamefont {Ziehe}, \citenamefont
  {Tkatchenko}, \citenamefont {von Lilienfeld},\ and\ \citenamefont
  {M{\"u}ller}}]{montavon2012learning}%
  \BibitemOpen
  \bibfield  {author} {\bibinfo {author} {\bibfnamefont {G.}~\bibnamefont
  {Montavon}}, \bibinfo {author} {\bibfnamefont {K.}~\bibnamefont {Hansen}},
  \bibinfo {author} {\bibfnamefont {S.}~\bibnamefont {Fazli}}, \bibinfo
  {author} {\bibfnamefont {M.}~\bibnamefont {Rupp}}, \bibinfo {author}
  {\bibfnamefont {F.}~\bibnamefont {Biegler}}, \bibinfo {author} {\bibfnamefont
  {A.}~\bibnamefont {Ziehe}}, \bibinfo {author} {\bibfnamefont
  {A.}~\bibnamefont {Tkatchenko}}, \bibinfo {author} {\bibfnamefont {O.~A.}\
  \bibnamefont {von Lilienfeld}}, \ and\ \bibinfo {author} {\bibfnamefont
  {K.-R.}\ \bibnamefont {M{\"u}ller}},\ }in\ \href@noop {} {\emph {\bibinfo
  {booktitle} {Advances in Neural Information Processing Systems}}}\ (\bibinfo
  {year} {2012})\ pp.\ \bibinfo {pages} {440--448}\BibitemShut {NoStop}%
\bibitem [{\citenamefont {von Lilienfeld}(2013)}]{von2013first}%
  \BibitemOpen
  \bibfield  {author} {\bibinfo {author} {\bibfnamefont {O.~A.}\ \bibnamefont
  {von Lilienfeld}},\ }\href@noop {} {\bibfield  {journal} {\bibinfo  {journal}
  {International Journal of Quantum Chemistry}\ }\textbf {\bibinfo {volume}
  {113}},\ \bibinfo {pages} {1676} (\bibinfo {year} {2013})}\BibitemShut
  {NoStop}%
\bibitem [{\citenamefont {Montavon}\ \emph {et~al.}(2013)\citenamefont
  {Montavon}, \citenamefont {Rupp}, \citenamefont {Gobre}, \citenamefont
  {Vazquez-Mayagoitia}, \citenamefont {Hansen}, \citenamefont {Tkatchenko},
  \citenamefont {M{\"u}ller},\ and\ \citenamefont {von
  Lilienfeld}}]{montavon2013machine}%
  \BibitemOpen
  \bibfield  {author} {\bibinfo {author} {\bibfnamefont {G.}~\bibnamefont
  {Montavon}}, \bibinfo {author} {\bibfnamefont {M.}~\bibnamefont {Rupp}},
  \bibinfo {author} {\bibfnamefont {V.}~\bibnamefont {Gobre}}, \bibinfo
  {author} {\bibfnamefont {A.}~\bibnamefont {Vazquez-Mayagoitia}}, \bibinfo
  {author} {\bibfnamefont {K.}~\bibnamefont {Hansen}}, \bibinfo {author}
  {\bibfnamefont {A.}~\bibnamefont {Tkatchenko}}, \bibinfo {author}
  {\bibfnamefont {K.-R.}\ \bibnamefont {M{\"u}ller}}, \ and\ \bibinfo {author}
  {\bibfnamefont {O.~A.}\ \bibnamefont {von Lilienfeld}},\ }\href@noop {}
  {\bibfield  {journal} {\bibinfo  {journal} {New J. Phys.}\ }\textbf {\bibinfo
  {volume} {15}},\ \bibinfo {pages} {095003} (\bibinfo {year}
  {2013})}\BibitemShut {NoStop}%
\bibitem [{\citenamefont {Hansen}\ \emph {et~al.}(2013)\citenamefont {Hansen},
  \citenamefont {Montavon}, \citenamefont {Biegler}, \citenamefont {Fazli},
  \citenamefont {Rupp}, \citenamefont {Scheffler}, \citenamefont {von
  Lilienfeld}, \citenamefont {Tkatchenko},\ and\ \citenamefont
  {M{\"u}ller}}]{hansen2013assessment}%
  \BibitemOpen
  \bibfield  {author} {\bibinfo {author} {\bibfnamefont {K.}~\bibnamefont
  {Hansen}}, \bibinfo {author} {\bibfnamefont {G.}~\bibnamefont {Montavon}},
  \bibinfo {author} {\bibfnamefont {F.}~\bibnamefont {Biegler}}, \bibinfo
  {author} {\bibfnamefont {S.}~\bibnamefont {Fazli}}, \bibinfo {author}
  {\bibfnamefont {M.}~\bibnamefont {Rupp}}, \bibinfo {author} {\bibfnamefont
  {M.}~\bibnamefont {Scheffler}}, \bibinfo {author} {\bibfnamefont {O.~A.}\
  \bibnamefont {von Lilienfeld}}, \bibinfo {author} {\bibfnamefont
  {A.}~\bibnamefont {Tkatchenko}}, \ and\ \bibinfo {author} {\bibfnamefont
  {K.-R.}\ \bibnamefont {M{\"u}ller}},\ }\href@noop {} {\bibfield  {journal}
  {\bibinfo  {journal} {J. Chem. Theory Comput.}\ }\textbf {\bibinfo {volume}
  {9}},\ \bibinfo {pages} {3404} (\bibinfo {year} {2013})}\BibitemShut
  {NoStop}%
\bibitem [{\citenamefont {Pozun}\ \emph {et~al.}(2012)\citenamefont {Pozun},
  \citenamefont {Hansen}, \citenamefont {Sheppard}, \citenamefont {Rupp},
  \citenamefont {M\"uller},\ and\ \citenamefont {Henkelman}}]{ML4Graeme2012}%
  \BibitemOpen
  \bibfield  {author} {\bibinfo {author} {\bibfnamefont {Z.~D.}\ \bibnamefont
  {Pozun}}, \bibinfo {author} {\bibfnamefont {K.}~\bibnamefont {Hansen}},
  \bibinfo {author} {\bibfnamefont {D.}~\bibnamefont {Sheppard}}, \bibinfo
  {author} {\bibfnamefont {M.}~\bibnamefont {Rupp}}, \bibinfo {author}
  {\bibfnamefont {K.-R.}\ \bibnamefont {M\"uller}}, \ and\ \bibinfo {author}
  {\bibfnamefont {G.}~\bibnamefont {Henkelman}},\ }\href@noop {} {\bibfield
  {journal} {\bibinfo  {journal} {J. Comp. Phys.}\ }\textbf {\bibinfo {volume}
  {136}},\ \bibinfo {pages} {174101} (\bibinfo {year} {2012})}\BibitemShut
  {NoStop}%
\bibitem [{\citenamefont {Snyder}\ \emph {et~al.}(2012)\citenamefont {Snyder},
  \citenamefont {Rupp}, \citenamefont {Hansen}, \citenamefont {M\"uller},\ and\
  \citenamefont {Burke}}]{ML4Kieron2012}%
  \BibitemOpen
  \bibfield  {author} {\bibinfo {author} {\bibfnamefont {J.~C.}\ \bibnamefont
  {Snyder}}, \bibinfo {author} {\bibfnamefont {M.}~\bibnamefont {Rupp}},
  \bibinfo {author} {\bibfnamefont {K.}~\bibnamefont {Hansen}}, \bibinfo
  {author} {\bibfnamefont {K.-R.}\ \bibnamefont {M\"uller}}, \ and\ \bibinfo
  {author} {\bibfnamefont {K.}~\bibnamefont {Burke}},\ }\href@noop {}
  {\bibfield  {journal} {\bibinfo  {journal} {Phys. Rev. Lett.}\ }\textbf
  {\bibinfo {volume} {108}},\ \bibinfo {pages} {253002} (\bibinfo {year}
  {2012})}\BibitemShut {NoStop}%
\bibitem [{\citenamefont {Sch\"utt}\ \emph {et~al.}(2014)\citenamefont
  {Sch\"utt}, \citenamefont {Glawe}, \citenamefont {Brockherde}, \citenamefont
  {Sanna}, \citenamefont {M\"uller},\ and\ \citenamefont
  {Gross}}]{GrossMLCrystals2014}%
  \BibitemOpen
  \bibfield  {author} {\bibinfo {author} {\bibfnamefont {K.~T.}\ \bibnamefont
  {Sch\"utt}}, \bibinfo {author} {\bibfnamefont {H.}~\bibnamefont {Glawe}},
  \bibinfo {author} {\bibfnamefont {F.}~\bibnamefont {Brockherde}}, \bibinfo
  {author} {\bibfnamefont {A.}~\bibnamefont {Sanna}}, \bibinfo {author}
  {\bibfnamefont {K.~R.}\ \bibnamefont {M\"uller}}, \ and\ \bibinfo {author}
  {\bibfnamefont {E.~K.~U.}\ \bibnamefont {Gross}},\ }\href {\doibase
  10.1103/PhysRevB.89.205118} {\bibfield  {journal} {\bibinfo  {journal} {Phys.
  Rev. B}\ }\textbf {\bibinfo {volume} {89}},\ \bibinfo {pages} {205118}
  (\bibinfo {year} {2014})}\BibitemShut {NoStop}%
\bibitem [{\citenamefont {Lopez-Bezanilla}\ and\ \citenamefont {von
  Lilienfeld}(2014)}]{lopez2014modeling}%
  \BibitemOpen
  \bibfield  {author} {\bibinfo {author} {\bibfnamefont {A.}~\bibnamefont
  {Lopez-Bezanilla}}\ and\ \bibinfo {author} {\bibfnamefont {O.~A.}\
  \bibnamefont {von Lilienfeld}},\ }\href@noop {} {\bibfield  {journal}
  {\bibinfo  {journal} {Phys. Rev. B}\ }\textbf {\bibinfo {volume} {89}},\
  \bibinfo {pages} {235411} (\bibinfo {year} {2014})}\BibitemShut {NoStop}%
\bibitem [{\citenamefont {Arsenault}\ \emph {et~al.}(2014)\citenamefont
  {Arsenault}, \citenamefont {Lopez-Bezanilla}, \citenamefont {von
  Lilienfeld},\ and\ \citenamefont {Millis}}]{arsenault2014machine}%
  \BibitemOpen
  \bibfield  {author} {\bibinfo {author} {\bibfnamefont {L.-F.}\ \bibnamefont
  {Arsenault}}, \bibinfo {author} {\bibfnamefont {A.}~\bibnamefont
  {Lopez-Bezanilla}}, \bibinfo {author} {\bibfnamefont {O.~A.}\ \bibnamefont
  {von Lilienfeld}}, \ and\ \bibinfo {author} {\bibfnamefont {A.~J.}\
  \bibnamefont {Millis}},\ }\href@noop {} {\bibfield  {journal} {\bibinfo
  {journal} {Phys. Rev. B}\ }\textbf {\bibinfo {volume} {90}},\ \bibinfo
  {pages} {155136} (\bibinfo {year} {2014})}\BibitemShut {NoStop}%
\bibitem [{\citenamefont {Ramakrishnan}\ \emph {et~al.}(2014)\citenamefont
  {Ramakrishnan}, \citenamefont {Dral}, \citenamefont {Rupp},\ and\
  \citenamefont {von Lilienfeld}}]{ramakrishnan2014quantum}%
  \BibitemOpen
  \bibfield  {author} {\bibinfo {author} {\bibfnamefont {R.}~\bibnamefont
  {Ramakrishnan}}, \bibinfo {author} {\bibfnamefont {P.~O.}\ \bibnamefont
  {Dral}}, \bibinfo {author} {\bibfnamefont {M.}~\bibnamefont {Rupp}}, \ and\
  \bibinfo {author} {\bibfnamefont {O.~A.}\ \bibnamefont {von Lilienfeld}},\
  }\href@noop {} {\bibfield  {journal} {\bibinfo  {journal} {Scientific Data}\
  }\textbf {\bibinfo {volume} {1}} (\bibinfo {year} {2014})}\BibitemShut
  {NoStop}%
\bibitem [{\citenamefont {Ramakrishnan}\ \emph {et~al.}(2015)\citenamefont
  {Ramakrishnan}, \citenamefont {Dral}, \citenamefont {Rupp},\ and\
  \citenamefont {von Lilienfeld}}]{ramakrishnan201bigdata}%
  \BibitemOpen
  \bibfield  {author} {\bibinfo {author} {\bibfnamefont {R.}~\bibnamefont
  {Ramakrishnan}}, \bibinfo {author} {\bibfnamefont {P.~O.}\ \bibnamefont
  {Dral}}, \bibinfo {author} {\bibfnamefont {M.}~\bibnamefont {Rupp}}, \ and\
  \bibinfo {author} {\bibfnamefont {O.~A.}\ \bibnamefont {von Lilienfeld}},\
  }\href@noop {} {\  (\bibinfo {year} {2015})},\ \bibinfo {note}
  {submitted}\BibitemShut {NoStop}%
\bibitem [{\citenamefont {Kung}(2014)}]{kung2014kernel}%
  \BibitemOpen
  \bibfield  {author} {\bibinfo {author} {\bibfnamefont {S.~Y.}\ \bibnamefont
  {Kung}},\ }\href@noop {} {\emph {\bibinfo {title} {Kernel Methods and Machine
  Learning}}}\ (\bibinfo  {publisher} {Cambridge University Press},\ \bibinfo
  {year} {2014})\BibitemShut {NoStop}%
\bibitem [{\citenamefont {Hansen}\ \emph {et~al.}(2015)\citenamefont {Hansen},
  \citenamefont {Biegler}, \citenamefont {Ramakrishnan}, \citenamefont {Pronobis}, 
  \citenamefont {von Lilienfeld}, \citenamefont {M\"uller},\ and\ \citenamefont {Tkatchenko}}]{BobPaper}%
  \BibitemOpen
  \bibfield  {author} {\bibinfo {author} {\bibfnamefont {K.}~\bibnamefont
  {Hansen}}, \bibinfo {author} {\bibfnamefont {F.}~\bibnamefont {Biegler}},
  \bibinfo {author} {\bibfnamefont {R.}\ \bibnamefont {Ramakrishnan}},
  \bibinfo {author} {\bibfnamefont {W.}\ \bibnamefont {Pronobis}},
  \bibinfo {author} {\bibfnamefont {O.~A.}\ \bibnamefont {von Lilienfeld}},
  \bibinfo {author} {\bibfnamefont {K.-R.}\ \bibnamefont {M\"uller}}, \ and\
  \bibinfo {author} {\bibfnamefont {A.}~\bibnamefont {Tkatchenko}},\
  }\href@noop {} {\  (\bibinfo {year} {2015})},\ \bibinfo {note}
  {submitted}\BibitemShut {NoStop}%
\bibitem [{\citenamefont {Ruddigkeit}\ \emph {et~al.}(2012)\citenamefont
  {Ruddigkeit}, \citenamefont {Van~Deursen}, \citenamefont {Blum},\ and\
  \citenamefont {Reymond}}]{ruddigkeit2012enumeration}%
  \BibitemOpen
  \bibfield  {author} {\bibinfo {author} {\bibfnamefont {L.}~\bibnamefont
  {Ruddigkeit}}, \bibinfo {author} {\bibfnamefont {R.}~\bibnamefont
  {Van~Deursen}}, \bibinfo {author} {\bibfnamefont {L.~C.}\ \bibnamefont
  {Blum}}, \ and\ \bibinfo {author} {\bibfnamefont {J.-L.}\ \bibnamefont
  {Reymond}},\ }\href@noop {} {\bibfield  {journal} {\bibinfo  {journal} {J.
  Chem. Inf. Model.}\ }\textbf {\bibinfo {volume} {52}},\ \bibinfo {pages}
  {2864} (\bibinfo {year} {2012})}\BibitemShut {NoStop}%
\bibitem [{\citenamefont {Johnson~III}(2013)}]{johnson2013nist}%
  \BibitemOpen
  \bibfield  {author} {\bibinfo {author} {\bibfnamefont {R.~D.}\ \bibnamefont
  {Johnson~III}},\ }\href@noop {} {\bibfield  {journal} {\bibinfo  {journal}
  {{\tt http://cccbdb.nist.gov}}\ } (\bibinfo {year} {2013})}\BibitemShut
  {NoStop}%
\bibitem [{\citenamefont {Tew}\ \emph {et~al.}(2007)\citenamefont {Tew},
  \citenamefont {Klopper}, \citenamefont {Heckert},\ and\ \citenamefont
  {Gauss}}]{tew2007basis}%
  \BibitemOpen
  \bibfield  {author} {\bibinfo {author} {\bibfnamefont {D.~P.}\ \bibnamefont
  {Tew}}, \bibinfo {author} {\bibfnamefont {W.}~\bibnamefont {Klopper}},
  \bibinfo {author} {\bibfnamefont {M.}~\bibnamefont {Heckert}}, \ and\
  \bibinfo {author} {\bibfnamefont {J.}~\bibnamefont {Gauss}},\ }\href@noop {}
  {\bibfield  {journal} {\bibinfo  {journal} {J. Phys. Chem. A}\ }\textbf
  {\bibinfo {volume} {111}},\ \bibinfo {pages} {11242} (\bibinfo {year}
  {2007})}\BibitemShut {NoStop}%
\bibitem [{\citenamefont {Hickey}\ and\ \citenamefont
  {Rowley}(2014)}]{hickey2014benchmarking}%
  \BibitemOpen
  \bibfield  {author} {\bibinfo {author} {\bibfnamefont {A.~L.}\ \bibnamefont
  {Hickey}}\ and\ \bibinfo {author} {\bibfnamefont {C.~N.}\ \bibnamefont
  {Rowley}},\ }\href@noop {} {\bibfield  {journal} {\bibinfo  {journal} {J.
  Phys. Chem. A}\ }\textbf {\bibinfo {volume} {118}},\ \bibinfo {pages} {3678}
  (\bibinfo {year} {2014})}\BibitemShut {NoStop}%
\bibitem [{\citenamefont {Hohm}(2000)}]{hohm2000there}%
  \BibitemOpen
  \bibfield  {author} {\bibinfo {author} {\bibfnamefont {U.}~\bibnamefont
  {Hohm}},\ }\href@noop {} {\bibfield  {journal} {\bibinfo  {journal} {J. Phys.
  Chem. A}\ }\textbf {\bibinfo {volume} {104}},\ \bibinfo {pages} {8418}
  (\bibinfo {year} {2000})}\BibitemShut {NoStop}%
\bibitem [{\citenamefont {Pearson}(1993)}]{pearson1993principle}%
  \BibitemOpen
  \bibfield  {author} {\bibinfo {author} {\bibfnamefont {R.~G.}\ \bibnamefont
  {Pearson}},\ }\href@noop {} {\bibfield  {journal} {\bibinfo  {journal} {Acc.
  Chem. Res.}\ }\textbf {\bibinfo {volume} {26}},\ \bibinfo {pages} {250}
  (\bibinfo {year} {1993})}\BibitemShut {NoStop}%
\bibitem [{dat()}]{datainvk}%
  \BibitemOpen
  \href@noop {} {}\bibinfo {howpublished}
  {For~data~contact~authors~or~visit\\{\tt
  http://tinyurl.com/qjkwxky}}\BibitemShut {NoStop}%
\end{thebibliography}%
\bibliographystyle{apsrev4-1}

\end{document}